\documentstyle[prl,eqsecnum,aps,epsf]{revtex}
\begin{document}
%\twocolumn[\hsize\textwidth\columnwidth\hsize\csname %
%@twocolumnfalse\endcsname
\draft
\title{
\large \bf
Analytical solutions to the third-harmonic generation in
trans-polyacetylene:\\
Application of dipole-dipole correlation on the single electron models
}
\author{Minzhong Xu}
\address{
Department of Chemistry, New York University, New York, NY 10003, U.S.A.
        }
\author{Xin Sun}
\address{ 
Department of Physics, Fudan University, and National Laboratory of Infrared
Physics, Shanghai 200433, P.R. China
}
\date{\today}
\maketitle
\bigskip
%{\center (Received: \hspace{13 cm})\\}
\begin{abstract}
The analytical solutions for the third-harmonic generation (THG) on infinite 
chains in both Su-Shrieffer-Heeger (SSH) and Takayama-Lin-Liu-Maki (TLM) models
of trans-polyacetylene are obtained through the scheme of dipole-dipole ($DD$)
correlation. They are not equivalent to the results obtained through
static current-current ($J_0J_0$) correlation or under polarization operator 
$\hat{P}$. The van Hove singularity disappears exactly in the analytical forms, 
showing that the experimentally observed two-photon absorption peak (TPA) in 
THG may not be directly explained by the single electron models.\\
\end{abstract}
%\phantom{.}
%]
%\narrowtext

%\pagebreak

\section{introduction}
The nonlinear optical (NLO) properties on conjugated polymers have always been 
received wide interest for study.\cite{messier,bredas0,heeger0,soos0}
Experimentally observed nonlinear phenomenon in conducting polymers, such as 
photoinduced absorption, bleaching, photoluminescence, ultrafast optical process
and exciton behaviors,\cite{lauchlan,orenstein,vardeny,hayes,harrison} 
have already encouraged the theorists for the possible 
explanations.\cite{gartstein,bredas1,kohler} 

In the early 90s, the large third order susceptibility ($\chi^{(3)}$) of 
trans-polyacetylene (PA) have been observed experimentally\cite{kajzar,heeger1} 
and have been received wide explanations 
theoretically.\cite{wwu,yuri,soos1,su1,su2,cwu1,cwu2,shuai,soos2,mazumdar} 
For a long time, the origin of the experimentally observed two-photon 
absorption peak (TPA)\cite{kajzar,heeger1} receives persistent discussions in
theory.\cite{wwu,yuri,soos1,su1,su2,cwu1,cwu2,shuai,soos2,mazumdar}
Based on the static current-current correlation ($J_0J_0$)\cite{wwu,yuri} and 
followed by Keldysh Green function methods\cite{keldysh}, TPA could be 
interpreted through the analytical forms under both SSH\cite{ssh} and TLM 
models\cite{tlm} for infinite chains. Besides 
the unnecessary application of nonequilibrium diagram techniques in a stable 
optical response problem,\cite{su2} however, the various numerical approaches 
from dipole formalism(${\bf E\cdot r}$), such as the results by Yu and 
Su\cite{su1,su2} from Butcher-Bloembergen-Shen dipole 
formula,\cite{butcher,bloembergen,shen} the results by Wu and 
Sun\cite{cwu1,cwu2} from Genkins-Mednis approach\cite{genkins,agrawal} and 
the results by Shuai and Br\'{e}das from sum-over-state(SOS) Orr-Ward
formalism,\cite{orr} have shown that the TPA is no longer existing if the 
damping in the energy is 
considered.\cite{su1,su2,cwu1,cwu2,shuai,soos2,mazumdar} 
Therefore, the two-photon cusp under the current schemes is considered to be a 
van Hove singularity\cite{gebhard} caused by the singular density of states 
(DOS) on the Fermi surface in one-dimensional ($1d$) systems. The fact of the 
nonresonant property in 
two-photon cusp is also noticed by Wu in his current formulations.\cite{wwu} 
However, besides the difficulty in explaining this TPA under current formulas, 
there exists another obvious difficulty in theory --- apparent zero frequency 
divergence (ZFD) in the definition of the current formalism.\cite{butcher} 
This problem actually was realized by Dakhnovskii and Pronin in their 
computations based on SSH model.\cite{yuri} Unfortunately, the results
of $\chi^{(3)}$ obtained from SSH model returns TLM model if the process of 
linearization procedures is made under $J_0J_0$ scheme.\cite{yuri} 

Because of the apparent divergence difficulties in current 
formulas\cite{butcher}, practically, intense studies based on dipole-dipole 
($DD$) correlation have been carried out to obtain the physical pictures of 
TPA in $\chi^{(3)}$ of trans-polyacetylene. On one hand, the weakly correlated 
and single electron theories
are applied in the numerical dipole computations, thus, the finite size
effect\cite{cwu2} and the lifetime assumption\cite{shuai} are seemingly good 
in generating a peak at exact position of TPA; On the other hand, the strongly 
correlated electron theories have also been applied to short chains, Soos and 
Ramesesha obtained the TPA based on Pariser-Parr-Pople (PPP) models, but it
is shifted too low in frequency and has too small an intensity\cite{soos1} 
relative to the experimental peak.\cite{kajzar} Recent 
work by Guo and Mazumdar shows that this TPA could be interpreted as a 
three-photon resonance based on extended Hubbard Hamiltonian.\cite{mazumdar} 
Whether or not the weakly correlated theory or even the single electron models 
such as SSH and TLM models are still suitable in describing TPA still remains 
as a question. Meanwhile, in obtaining the qualitatively correct result, it was 
suggested by Etemad and Soos that the attention should be made to the
$\chi^{(3)}$ frequency dependences rather than magnitudes.\cite{soos3} It is
certainly reasonable in physical point of view since the experiments are more
sensitive to detect frequency dependences. Thus, some typical approximations 
are made in the dipole formula. 
Substitution of the dipole operator $\hat{D}$ by the polarization operator 
$\hat{P}$, namely, the dipole approximation\cite{soos2,mukamel1} has been 
extensively applied in the nonlinear optical formula, however, $\chi^{(1)}$ 
computations have shown both quantitatively and qualitatively different results 
between $\hat{P}$ and dipole operator $\hat{D}$.\cite{mxu} Whether the problem 
caused by the dipole approximation still existing in $\chi^{(3)}$ is still 
under our concern. Furthermore, the static dipole 
formulas\cite{soos1,su1,su2,cwu1,cwu2,shuai,soos2,mazumdar} and static 
current formulas\cite{wwu,butcher} are conventionally considered to be 
equivalent.\cite{su1,butcher,bloembergen,shen} However, as we pointed out in a 
recent work,\cite{mxu} it has ZFD problem in the static current formalism when 
the gauge phase factor is ignored, the static formalism under two gauges should 
not return the same results under the exact same basis of unperturbed wave 
functions.\cite{mxu} Whether this nonequivalence between $J_0J_0$ and $DD$
correlations existing in the specific case of $\chi^{(3)}$ of 
trans-polyacetylene is another interest topic. Although the qualitative 
features have already been reported in the numerical solutions based on $DD$ 
correlation,\cite{soos1,su1,su2,cwu1,cwu2,shuai,soos2,mazumdar}
the exact analytical solutions from the dipole formulas of $\chi^{(3)}$ 
have not been obtained previously. The analytical results, if obtained,
will be very helpful for us to directly check all above concerns and give more
information.

Fortunately, both SSH and TLM models are single electron models and can be
exactly
solvable for the nonlinear susceptibilities under perturbated schemes, although
the efforts will be no doubt paid to cumbersome computations. After 
choosing the long wavelength approximation which are conventionally used 
previously,\cite{wwu,yuri,soos1,su1,su2,cwu1,cwu2,shuai,soos2,mazumdar,mxu} 
we could directly compared those analytical results 
between two gauges for both models.
The paper is organized as follows, we will first discuss the SSH Hamiltonian 
in Sec.II.A. The analytical form of $\chi^{(3)}$ by the dipole-dipole (DD) 
correlation is obtained and compared with the analytical form by $J_0J_0$ 
correlation for SSH models in Sec.II.B. In Sec.II.C, we will obtain the
results under the polarization operator $\hat{P}$, the qualitative and 
quantitative difference between the results under $\hat{D}$ and $\hat{P}$ are 
outlined. In Sec.III, we will obtain the exact analytical forms under 
$DD$ correlation for TLM models and directly compared with those under 
$J_0J_0$ correlation. The discussions
on the possible implications of this work are made in Sec.IV and a brief 
conclusion is made in Sec.V.

\section{Third order susceptibilities for SSH models under dipole formula}
\subsection{SSH Hamiltonian in real and momentum spaces}
Based on periodic tight-binding-approximations, The SSH Hamiltonian\cite{ssh} 
is given by:
\begin{eqnarray}
H_{SSH}=-\sum_{l,s} \left[ t_0+(-1)^l \frac{\Delta}{2} \right]
(\hat{C}_{l+1,s}^{\dag}\hat{C}_{l,s}^{}+\hat{C}_{l,s}^{\dag}\hat{C}_{l+1,s})^{},
\label{Hssh}
\end{eqnarray}
where $t_0$ is the transfer integral between the nearest-neighbor sites,
$\Delta$ is the gap parameter and $\hat{C}_{l,s}^{\dag}(\hat{C}_{l,s})$
creates(annihilates) an $\pi$ electron at site $l$ with spin $s$. In
continuum limitation, above SSH model will give the TLM model.\cite{tlm}
For the SSH model, each site is occupied by one electron. 

Under the $DD$ correlation, the interaction Hamiltonian is expressed by
$\hat{H}_{\bf E\cdot r}=-e{\bf E\cdot r}=-{\bf D\cdot E}$,
To provide the periodic property and to avoid the ill-definition of position
operator $\hat{\bf r}$, we should express the position operator $\hat{\bf r}$
under the Bloch states: $\displaystyle |n, {\bf k}> = u_{n, {\bf k}}
({\bf r})e^{i{\bf k \cdot r}}$, where $n$ and ${\bf k}$ are band index and
crystal momentum correspondingly, $u_{n, {\bf k}}({\bf r})$ is the periodic
function under the translation of lattice vector.\cite{callaway} Thus, we 
obtain:
\begin{eqnarray}
{\bf r}_{n {\bf k}, n' {\bf k'}}= i \delta_{n,n'}{\bf \nabla_{k}}
\delta({\bf k}-{\bf k'}) + \Omega_{n,n'}({\bf k})\delta({\bf k}-{\bf k'}),
\label{r}
\end{eqnarray}
and
\begin{eqnarray}
\displaystyle \Omega_{n,n'}({\bf k})=\frac{i}{v}\int_{v}
u_{n,{\bf k}}^*({\bf r}){\bf \nabla_{k}} u_{n', {\bf k}}({\bf r}) d {\bf r},
\end{eqnarray}
where $v$ is unit cell volume.

We change Hamiltonian Eq.(\ref{Hssh}) into the momentum space by applying
the following consecutive transformations:
\begin{equation}
\left \{
\begin{array}{l}
\displaystyle \hat{C}_{l_o,s}=\frac{1}{\sqrt{N}}
\sum_{-\frac{\pi}{2a}\le k \le \frac{\pi}{2a}}
(\hat{C}^v_{k,s}+\hat{C}^c_{k,s})e^{ikR_{l_o}},\\
\\
\displaystyle  \hat{C}_{l_e,s}=\frac{1}{\sqrt{N}}
\sum_{-\frac{\pi}{2a}\le k \le \frac{\pi}{2a}}
(\hat{C}^v_{k,s}-\hat{C}^c_{k,s})e^{ikR_{l_e}},
\end{array}
\right.
\label{trk1}
\end{equation}
and
\begin{equation}
\left \{
\begin{array}{l}
\displaystyle \hat{a}^v_{k,s}=-i\gamma_k\hat{C}^v_{k,s}+\xi_k\hat{C}^c_{k,s},\\
\\
\displaystyle \hat{a}^c_{k,s}=i\xi_k\hat{C}^v_{k,s}+\gamma_k\hat{C}^c_{k,s},
\end{array}
\right.
\label{trk2}
\end{equation}
with
\begin{equation}
\left \{
\begin{array}{l}
\displaystyle \gamma_k=\frac{1}{\sqrt{2}}\sqrt{1+\frac{2t_0 cos(ka)}
{\varepsilon(k)}},\\
\\
\displaystyle \xi_k=\frac{sgn(k)}{\sqrt{2}}\sqrt{1-\frac{2t_0 cos(ka)}
{\varepsilon(k)}},
\end{array}
\right.
\end{equation}
where
\begin{eqnarray}
\varepsilon (k)= \sqrt{\left[ 2 t_0 cos(ka) \right]^2+\left[ \Delta sin(ka)
\right]^2},
\label{ek}
\end{eqnarray}
and $\hat{a}^{\dag c}_{k,s}(t)$ and $\hat{a}^{\dag v}_{k,s}(t)$ are
the excitations of electrons in the conduction band and the valence
band with momentum $k$ and spin $s$,
$R_{l_o}$ and $R_{l_e}$ are odd and even position defined by
\begin{eqnarray}
R_l=la+(-1)^l u,
\label{rl}
\end{eqnarray}

If we choose the spinor description $\hat{\psi}_{k,s}^{\dag}(t)$=
$(\hat{a}^{\dag c}_{k,s}(t)$, $\hat{a}^{\dag v}_{k,s}(t))$ and apply the long
wavelength approximation,\cite{butcher} SSH Hamiltonian including
${\bf E \cdot r}$ in momentum space is described by:
\begin{eqnarray}
\hat{H}_{SSH}(k,t)= \hat{H}_0+\hat{H}_{\bf E \cdot r},
\end{eqnarray}
where
\begin{eqnarray}
\hat{H}_0=\sum_{-\frac{\pi}{2a}\le k\le\frac{\pi}{2a},s}
\varepsilon(k) \hat{\psi}_{k,s}^{\dag}(t) \sigma_{3}
\hat{\psi}_{k,s}(t)
\label{Hsshk}
\end{eqnarray}
and
\begin{eqnarray}
\hat{H}_{\bf E \cdot r}=- \hat{D} \cdot E_0 e^{i\omega t}.
\end{eqnarray}
By Eq. (\ref{r}), the dipole operator $\hat{D}$ could be obtained
as follows:\cite{mxu}
\begin{eqnarray}
\hat{D}= e \sum_{-\frac{\pi}{2a}\le k\le\frac{\pi}{2a},s}
(\beta(k)\, \hat{\psi}_{k,s}^{\dag}
\sigma_{2}\hat{\psi}_{k,s}
 +i \frac{\partial}{\partial k} \, \hat{\psi}_{k,s}^{\dag}\hat{\psi}_{k,s}),
\label{D}
\end{eqnarray}
where
\begin{eqnarray}
\beta(k)=-\displaystyle \frac{\Delta t_0 a }{ \varepsilon^2(k)},
\end{eqnarray}
is the coefficient related to the interband transition between the conduction
and valence bands in a unit cell $2a$ and the second term in Eq.(\ref{D}) is
related to the intraband transition,\cite{mxu} $e$ is the electric charge and
$\vec{\sigma}$ are the Pauli matrixes. We have already omitted the relative
distortion $\eta$($\equiv 2u/a$) in this computation since the contribution
of $\eta$ is quite small according to the linear case.\cite{mxu}

\subsection{Analytical form of $\chi^{(3)}$ through $DD$ correlation}
Within the semiclassical theory of radiation,\cite{butcher,bloembergen,shen}
the electrical field is treated classically and the third order susceptibility
$\chi^{(3)}$ is described by:\cite{butcher,bloembergen,shen} 
\begin{eqnarray}
&\chi&^{(3)}(\Omega; \omega_1, \omega_2, \omega_3)= \frac{1}{3!}
\left[ \frac{i}{\hbar} \right]^3 \int  d{\bf r}_1 d{\bf r}_2 d{\bf r}_3
\int dt_1 dt_2 dt_3 \nonumber \\
& &\int d{\bf r} dt\, e^{-i {\bf k \cdot r}+ i \Omega t} \langle \hat{T}
\hat{{\bf D}} ({\bf r},t) \hat{{\bf D}}({\bf r}_1,t_1)
\hat{{\bf D}}({\bf r}_2,t_2) \hat{{\bf D}} ({\bf r}_3,t_3) \rangle,
\label{DD}
\end{eqnarray}
where $\displaystyle \Omega \equiv -\sum_{i=1}^{3} \omega_{i}$ and T is the 
time-ordering operator with $\hat{{\bf D}}$ is a dipole operator.

The third-harmonic generation (THG) is defined by setting 
$\omega_1=\omega_2=\omega_3$.
Following similiar procedures as we have done in the linear 
computations,\cite{mxu} by Eq.(\ref{D}) and Eq.(\ref{DD}), we obtained 
the following expression for $\chi_{SSH}^{THG}(\omega_1)$$\equiv$
$\chi_{SSH}^{(3)}(-3\omega_1,\omega_1,\omega_1,\omega_1)$:
\begin{eqnarray}
\chi_{SSH}^{THG}(\omega_1)= \frac{2 e^4 n_0}{\hbar^3} \sum_k \int 
\frac{d\omega}{2\pi}\Biggl\{ (&\beta&(k)\sigma_2+i\frac{\partial}{\partial k})
G(k,\omega) \nonumber \\
(&\beta&(k)\sigma_2+i\frac{\partial}{\partial k}) G(k,\omega-\omega_1)
\nonumber \\ 
(&\beta&(k)\sigma_2+i\frac{\partial}{\partial k}) G(k,\omega-2\omega_1) 
\nonumber \\
(&\beta&(k)\sigma_2+i\frac{\partial}{\partial k}) G(k,\omega-3\omega_1)
\Biggr\},
\label{ssh0}
\end{eqnarray}
where $\beta(k)$ is defined in Eq.(\ref{D}) and $n_0$ is the number of chains
in unit cross area and the polymer chains are assumed to be oriented,
and Green Function $G(k,\omega)$ defined as following:\cite{mxu} 
\begin{eqnarray}
\displaystyle G(k,\omega)=
\frac{\omega+\omega_k\sigma_3}{\omega^2-\omega^2_k+i\epsilon},
\label{green}
\end{eqnarray}
with $\omega_k \equiv \varepsilon(k) / \hbar \text{ and } \epsilon \equiv 0^+$.

After tedious derivation, we obtained the following analytical expressions 
for the third-harmonic generation for the SSH models 
$\chi_{SSH}^{THG}(\omega)$:
\begin{eqnarray}
\chi_{SSH}^{THG}(\omega)= \frac{2 e^4 n_0}{\hbar^3} \sum_k \Biggl\{
&+&\frac{\beta^4(k)}{2 \omega(2\omega_k+\omega)(2\omega_k+3\omega)}
\nonumber \\
&-&\frac{\beta^4(k)}{2 \omega(2\omega_k-\omega)(2\omega_k-3\omega)}
\nonumber \\
&-&\frac{\beta(k)}{(2\omega_k+3\omega)}\frac{\partial}{\partial k}\left[
   \frac{1}{(2\omega_k+2\omega)}\frac{\partial}{\partial k} \left[
   \frac{\beta(k)}{(2\omega_k+\omega)} \right] \right] \nonumber \\
&-&\frac{\beta(k)}{(2\omega_k-3\omega)}\frac{\partial}{\partial k}\left[
   \frac{1}{(2\omega_k-2\omega)}\frac{\partial}{\partial k} \left[
   \frac{\beta(k)}{(2\omega_k-\omega)} \right] \right]
\Biggr\},
\label{ssh}
\end{eqnarray}
The above expression Eq.(\ref{ssh}) are almost identical to Eq.(A13) in Agrawal 
{\it et al.}'s work\cite{agrawal} except one more term (last term in (A13)), 
and with the following substitution is applied in Eq.(A13):
\begin{equation} 
\left\{
\begin{array}{l}
\Omega_{vv}=\Omega_{cc}=0,\\
\\
\Omega_{vc}=-\Omega_{cv}=\beta(k)
\end{array}
\right.
\label{st}
\end{equation}
Eq.(\ref{ssh}) is also quite similar to Eq.(11) in Wu and Sun's 
result\cite{cwu1} based on Genkins-Mednis approach\cite{genkins} with the
above substitution (\ref{st}) made and if the symmetric procedures on the 
negative frequency term should be applied.

For infinite chains, we could separate two full derivative terms of $k$ from 
Eq.(\ref{ssh}) as follows:
\begin{eqnarray}
I_T=\frac{2e^4n_0}{\hbar^3} \sum_k \Biggl\{
&-&\frac{\partial}{\partial k}\left[\frac{\beta(k)}{(2\omega_k+3\omega)
(2\omega_k+2\omega)}\frac{\partial}{\partial k} \left[
   \frac{\beta(k)}{(2\omega_k+\omega)} \right] \right] \nonumber \\
&-&\frac{\partial}{\partial k}\left[\frac{\beta(k)}{(2\omega_k-3\omega)
(2\omega_k-2\omega)}\frac{\partial}{\partial k} \left[
   \frac{\beta(k)}{(2\omega_k-\omega)} \right] \right]
\Biggr\},
\end{eqnarray}
If we consider following two facts in actual optical process: (i). The 
velocity on the Fermi surface $\nabla_k\varepsilon (k)|_{k=\pm 
\frac{\pi}{2a}}=0$. (ii). The life time of the states, thus, we should include 
the damping in the process, that is, $\omega_k\to\omega_k+ i\epsilon_k$ and 
$\epsilon_k\ne0$. The above full 
derivative term $I_T$ after integral of $k$ then is 0 for the infinite chains.
Eq.(\ref{ssh}) can be simplified as following:
\begin{eqnarray}
\chi_{SSH}^{THG}(\omega)
&=&\chi_0^{(3)} \frac{45}{128} \int_1^{\frac{1}{\delta}}
\frac{d x}{ [(1-\delta^2 x^2)(x^2-1)]^{\frac{1}{2}}} 
\Biggl\{\frac{37-24(1+\delta^2)x^2+12\delta^2x^4}{8 x^8(x^2-z^2)} \nonumber \\
&+&\frac{9\left[ 243-216(1+\delta^2)x^2+188\delta^2 x^4 \right]}
{8 x^8(x^2-(3 z)^2)} \Biggr\} \nonumber \\
&=&\chi_0^{(3)} \frac{5}{1024z^8}\biggl\{
-336E(\frac{\pi}{2},\sqrt{1-\delta^2})
+120z^2\delta^2F(\frac{\pi}{2},\sqrt{1-\delta^2}) \nonumber \\
&+&\frac{8z^4}{5}\left[(-12+7\delta^2-12\delta^4)E(\frac{\pi}{2},
\sqrt{1-\delta^2})+6(1+\delta^2)\delta^2F(\frac{\pi}{2},\sqrt{1-\delta^2})
\right] \nonumber \\
&+&9(37-24(1+\delta^2)z^2+12\delta^2z^4)g(z)
+(3-24(1+\delta^2)z^2+188\delta^2z^4)g(3z) \biggr\},
\label{sshdd}
\end{eqnarray}
and
\begin{eqnarray}
g(mz)=\frac{n_m \delta}{\delta^2-1}\Pi(\frac{\pi}{2},n_m,\sqrt{1-\delta^2}),
\quad n_m\equiv\frac{1-\delta^2}{(\delta mz)^2-1}.
\label{gz}
\end{eqnarray}
where $\displaystyle \chi_0^{(3)} \equiv \frac{8}{45}\frac{e^4 n_0}{\pi}
\frac{(2 t_0 a)^3}{\Delta^6}$,  $ x \equiv \hbar \omega_k / \Delta$, 
$z \equiv \hbar \omega /(2\Delta)$ and $\delta \equiv \Delta /(2 t_0)$.
$F, E, \Pi$ are first, second and third kind of complete elliptical 
integrals.\cite{table} There is exactly no two photon cusp in 
Eq.(\ref{sshdd}). This result is consistent with Yu and Su, Shuai and 
Br\`{e}das' results. And it is quite different from the expression obtained 
from $J_0J_0$ correlation, which shows the following form for 
$\chi^{THG}_{SSHjj}$ of SSH models:\cite{yuri}
\begin{eqnarray}
\chi^{THG}_{SSHjj}=&B&z^{-8}\{[5-8z^2(1+\delta^2)+20z^4\delta^2]g(z)
-8[1-4z^2(1+\delta^2)+16z^4\delta^2]g(2z) \nonumber \\
&+&[3-24z^2(1+\delta^2)+188z^4\delta^2]g(3z)
-8\delta^2z^4E(\frac{\pi}{2},\sqrt{1-\delta^2})\},
\label{sshjj}
\end{eqnarray}
where $B\equiv 5\chi^{(3)}_0/(1024z^8)$, same as the constant as defined in 
Eq.(\ref{sshdd}) and g(z) is defined by Eq.(\ref{gz}). We should point out
that the elliptical form of Eq.(\ref{sshjj}) \cite{yuri} is the exact the same 
as the integral form of Eq.(11) in Wu and Sun's result,\cite{cwu1} which was 
derived from the incomplete Genkins-Mednis approach.\cite{genkins,mxu} The
disappearance of ZFD under static current schemes is somewhat puzzling 
or simply should be understood as coincidence for the specific SSH model.
As we pointed out in a recent work,\cite{mxu} the ignorance of gauge phase 
factor in the static current formula will cause ZFD in the susceptibilities 
and non-equivalent result from the dipole formula. The derivation in the 
previous current work\cite{yuri} does not consider the gauge phase factor's 
effect, it will cause ZFD even for linear case if following a straightforward
computations.\cite{mxu} 

Obviously, we observe the qualitatively difference for results under $DD$ 
(Eq.\ref{sshdd})) and $J_0J_0$ (Eq.(\ref{sshjj})) correlation  especially 
for $z=1/2$ and $z=1$. 
The inequivalence could be understood by the ignore of the 
gauge phase factor in optical response theory.\cite{mxu}

Let $x\to x+i\epsilon$ in the integral of Eq.(\ref{sshdd}), the comparison graph
of absolute values between Eq.(\ref{sshdd}) and Eq.(\ref{sshjj}) for the SSH
models are plotted in Fig.1 with the following parameters chosen for 
trans-polyacetylene:\cite{su1,su2,cwu1,cwu2} $t_0=2.5 eV$, $\Delta=0.9 eV$, 
$n_0=3.2 \times 10^{14} cm^{-2}$, $a=1.22 \AA$ and $\epsilon \sim 0.03$, we 
have $\delta=0.18$ and $\chi_0^{(3)} \approx 1.0 \times 10^{-10}$ esu.

\begin{figure}
\vskip -10pt
\centerline{
\epsfxsize=7cm \epsfbox{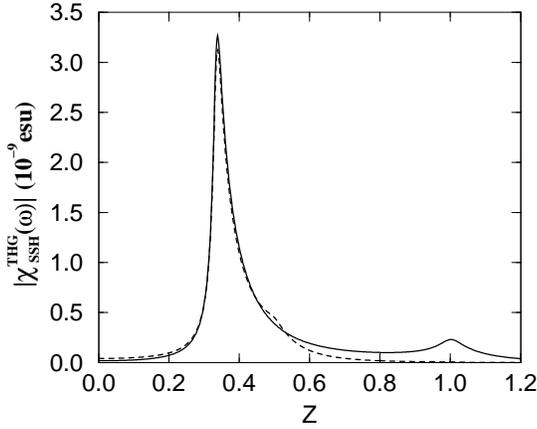}
}
\vskip 0pt
\caption{Computed $DD$ values (solid line) vs. $J_0J_0$ values (dashed line)
of $|\chi_{SSH}^{THG}(\omega)|$ with $z \equiv \hbar\omega/(2\Delta)$ and
$\epsilon=0.03$.}
\end{figure}

From the graph, we find another very obvious peak at $z=1$ at a ratio of $1/10$
of the peak at $z=1/3$, which is also different comparing with the previous 
theoretical
computations.\cite{wwu,yuri,soos1,su1,su2,cwu1,cwu2,shuai,soos2,mazumdar}
However, this peak has not been reported by the experiments because
it is out of the scanning range of photon energy.\cite{kajzar,heeger1}

\subsection{Results under polarization operator $\hat{P}$}
In a recent work, we discussed the different results caused by polarization
operator $\hat{P}$ in linear response\cite{mxu}. The polarization operator 
$\hat{P}$ is extensively applied in the NLO theory.\cite{soos2,mukamel1} 
Based on SSH model, we can do a comparison of $\chi^{THG}_{SSH}$ between 
$\hat{D}$ and $\hat{P}$. 

The polarization operator $\hat{P}$ is defined in real space as:
\begin{eqnarray}
\hat{P}= \sum_{l} R_{l}\hat{C}^{\dagger}_{l} \hat{C}_{l}.
\label{P}
\end{eqnarray}

If the chain region $l$ is going from $1$ to $N$, we obtain the
unit polarization $\hat{P}^{unit}(k)$ from the total polarization
$\hat{P}^{total}(k)$ in the momentum space:
\begin{eqnarray}
\hat{P}^{unit}(k)=\lim_{N\to\infty}\displaystyle\frac{\hat{P}^{total}(k)}{N}=
\frac{ea}{2}\sum_{k,s} \psi^{\dag}_{k,s}\sigma_2\psi_{k,s}
\end{eqnarray}
Substituting $\hat{D}$ with $\hat{P}$ in Eq.(\ref{DD}), we obtain the
$\chi^{THG}_{SSH_P}$ for SSH models as following for infinite chains:
\begin{eqnarray}
\chi^{THG}_{SSH_P}=\chi_0^{(3)}\frac{45\delta^4}{128}
\int_1^{\frac{1}{\delta}}\frac{d x}{ [(1-\delta^2 x^2)(x^2-1)]^{\frac{1}{2}}}
\Biggl\{\frac{1}{x^2-z^2}-\frac{9}{x^2-(3z)^2}\Biggr\}
\end{eqnarray}
The above expressions lacks TPA. However, the magnitude is too small compared
with results under $\hat{D}$. ($|\chi^{THG}_{SSH_P}|$ from $\hat{P}$ is about 
$10^{-4}$ of $|\chi^{THG}_{SSH}|$ from $\hat{D}$ if $\delta=0.18$ for 
trans-polyacetylene.) Another peak
at $z=1/(3\delta)\approx 1.85$ will shown up through ${\bf P\cdot E}$, this 
peak corresponding to the transition from the bottom of valence band to the 
top of conduction band. This peak in infinite chains seems not to agree with
`unklapp enhancement' in the solid states.\cite{gebhard} While in $DD$ 
computations, this peak at $z=1/(3\delta)$ will not be very obvious. The 
comparison on the absolute value between $\chi^{THG}_{SSH_P}$ and 
$\chi^{THG}_{SSH}$ will be shown in Fig.2. From the graph, the different 
shapes of curves are quite
obvious. Besides the possible ill-definition of polarization operator 
$\hat{P}$,\cite{gebhard,mxu,callaway} both the qualitative and quantitative 
features obtained from the $\hat{P}$ are quite different even for nonlinear
optical response theory. Thus it will not be quite suitable in using $\hat{P}$
to obtain the equivalent results as $\hat{D}$.

\begin{figure}
\vskip -10pt
\centerline{
\epsfxsize=7cm \epsfbox{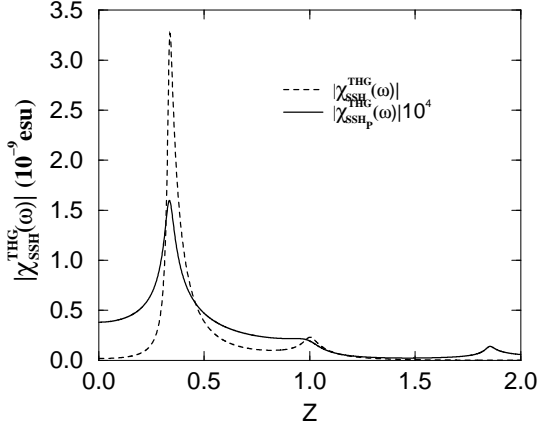}
}
\vskip 0pt
\caption{Computed $|\chi_{SSH_P}^{THG}(\omega)|$ under polarization operator 
$\hat{P}$ (solid line) vs. computed $|\chi_{SSH}^{THG}(\omega)|$ (dashed line)
under dipole operator $\hat{D}$ with $z \equiv \hbar\omega/(2\Delta)$,
$\epsilon=0.03$ and a magnification of $10^4$ in 
$|\chi_{SSH_P}^{THG}(\omega)|$.}
\end{figure}
 
\section{Third order susceptibility for TLM model under dipole formula}
The results under TLM model for the $DD$ correlation can be obtained by 
setting $\delta \to 0$ and $(2t_0a)\to \hbar v_F$ in 
Eq.(\ref{sshdd}).\cite{mxu} Since we have the following 
properties for the complete elliptical integrals:\cite{table}
\begin{equation}
\left\{
\begin{array}{l}
\displaystyle \lim_{\delta \to 0} E(\frac{\pi}{2},\sqrt{1-\delta^2})=1,\\
\\
\displaystyle \lim_{\delta \to 0} \delta^2 
F(\frac{\pi}{2},\sqrt{1-\delta^2})=0,\\
\\
\displaystyle \lim_{\delta \to 0} 
\Pi(\frac{\pi}{2},n_m,\sqrt{1-\delta^2})=f(mz),
\end{array}
\right.
\end{equation}
where
\begin{equation}
f(z) \equiv \left \{
\begin{array}{lr}
\displaystyle  {\arcsin (z)\over z \sqrt{1-z^2}}  &(z^2<1),\\
\\
\displaystyle  -{\cosh^{-1} (z)\over z\sqrt{z^2-1}}+\displaystyle
{i\pi \over 2 z\sqrt{z^2-1}} &\ \ (z^2>1).
\end{array}
\right.
\label{fz}
\end{equation}

We will obtain the following expressions for THG under TLM models for $DD$
correlation,
\begin{eqnarray}
\chi_{TLM}^{THG}(\omega)= \chi_0^{(3)} \frac{5}{1024z^8} \, \Biggl\{
-336-\frac{96z^4}{5}
+9(37-24 z^2)f(z)+3(1-8 z^2)f(3z) \Biggr\}
\label{tlmdd}
\end{eqnarray}
where $\chi_0^{(3)}$ defined in Eq.(\ref{sshdd}). The $\chi_{TLM}^{THG}$ is 
plotted in Fig.3. And no singular property such as van Hove singular shows 
in $\chi^{THG}_{TLM}$ under $DD$ correlation.

In $J_0J_0$ correlation, the result of $\chi^{THG}_{TLMjj}$ obtained by Wu as
follows:\cite{wwu}
\begin{eqnarray}
\chi_{TLMjj}^{THG}(\omega)= \chi_0^{(3)} \frac{5}{1024z^8} \{
(5-8z^2)f(z)-8(1-4z^2)f(2z)+3(1-8z^2)f(3z)\}
\label{tlmjj}
\end{eqnarray}

The comparison between our result($DD$) and Wu's result 
($J_0J_0$)\cite{wwu} on absolute value of $\chi_{TLM}^{THG}$ is
plotted in Fig.4. If the zero frequency limitation is made, let $z\to0$,
we obtain the different values for the zero frequency limit for both $DD$ and
$J_0J_0$ correlations:
\begin{equation}
\left\{
\begin{array}{ll}
\chi_{TLM}^{THG}(0)= \displaystyle \frac{5}{28}\chi_0^{(3)} \approx 1.8 \times
10^{-11} esu. \quad
\text{$DD$ correlation} \\
\\
\chi_{TLMjj}^{THG}(0)= \displaystyle \frac{1}{2}\chi_0^{(3)} \approx 
5.0 \times 10^{-11} esu. \quad \text{$J_0J_0$
correlation}\\
\end{array}
\right.
\end{equation}

The nonequivalent results between $DD$ and $J_0J_0$ correlation are quite
obvious even for the static limit. The disappearance of ZFD could be 
understood as coincidence.\cite{mxu}

\begin{figure}
\vskip -10pt
\centerline{
\epsfxsize=7cm \epsfbox{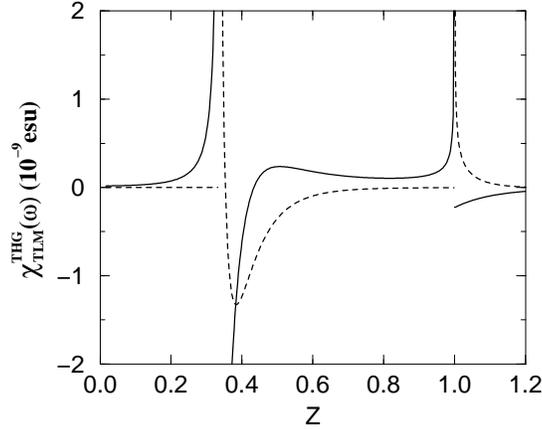}
}
\vskip 0pt
\caption{The real part (solid line) and the imaginary part (dashed line)
of $\chi_{TLM}^{THG}(\omega)$ with $z \equiv \hbar\omega/(2\Delta)$.}
\end{figure}

\begin{figure}
\vskip -10pt
\centerline{
\epsfxsize=7cm \epsfbox{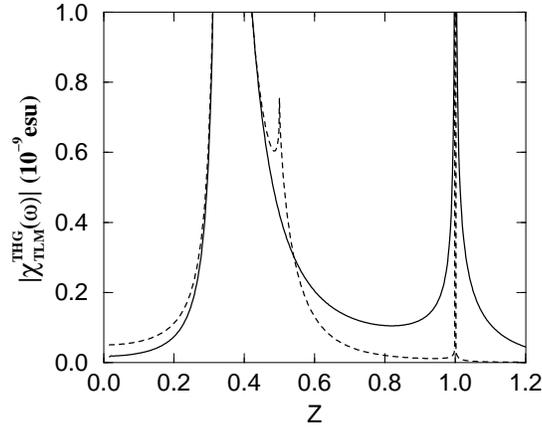}
}
\vskip 0pt
\caption{Computed $DD$ values (solid line) vs. $J_0J_0$ values (dashed line)
of $|\chi_{TLM}^{THG}(\omega)|$ with $z \equiv \hbar\omega/(2\Delta)$.}
\end{figure}

\section{discussions}
The analytical forms Eq.(\ref{sshdd}) and Eq.(\ref{tlmdd}) show that there
will be exactly no two-photon peak or even no van Hove singularity under $DD$
correlation. They are qualitatively different from the expressions 
Eq.(\ref{sshjj}) and Eq.(\ref{tlmjj}) derived from $J_0J_0$ correlation.
The results with no TPA peak under single electron models are certainly 
reasonable in physical pictures, which is also agrees with the previous
arguments on this problem.\cite{soos1,su1,su2,cwu1,cwu2,shuai,soos2,mazumdar} 
The nonequivalence between two gauges also provided some evidence that the
gauge phase factor in the optical response theory can not be 
ignored, and the static current expressions\cite{wwu,yuri,butcher} can not
be directly used and return the exact same results as $DD$ correlation.
On one hand, $DD$ correlation should be much more suitable to 
study the nonlinear susceptibilities than $JJ$ correlation.\cite{mxu}
On the other hand, based on the exact expression of dipole operator $\hat{D}$,
we find that polarization operator $\hat{P}$ can not be directly applied in
periodic systems where the electrons are certainly not bounded. And $\hat{P}$ 
will returns the qualitatively different curves from the rigorous results.
From the results we obtained in this paper, it seems that the direct 
application of single electron models may not be suitable in explaining for 
TPA. It also implies that the size effect\cite{cwu2}, lifetime 
considerations\cite{soos1,su1,su2,cwu1,cwu2,shuai,soos2,mazumdar}, exciton 
effects\cite{soos0,soos1,mazumdar}, quantum fluctuations or disorders should 
be included to explain this non-trivial TPA. Both SSH and TLM models still can
be served as the good basis to include all those interactions. 
  
\section{conclusions}
The analytical solutions to the THG for both SSH and TLM models obtained
under the $DD$ correlation computations, showing nonexistence of the 
two-photon peak exactly. The nonequivalent results between $DD$ and $J_0J_0$ 
correlations provide an evidence for the possible explanations through gauge 
phase factor.\cite{mxu} In actual practical computations, $DD$ will be much 
more suitable for obtaining reasonable results than $J_0J_0$ correlation, 
however, the replacement of $\hat{D}$ by the polarization operator $\hat{P}$ 
will result in qualitatively different results especially in periodic systems. 

\acknowledgments 
Very helpful discussions with Professor J.L. Birman, Professor Y.R. Shen, 
Professor Z.G. Shuai, Professor Z.G. Soos, Professor C.Q. Wu and Dr. Z.G. Yu 
are acknowledged.
This work was supported by Chemistry Department, New York University, 
the Project $863$ and the National Natural Science Foundation of China
(59790050, 19874014).

\end{document}